\documentclass[]{interact}

\usepackage{epstopdf}% To incorporate .eps illustrations using PDFLaTeX, etc.
\usepackage[caption=false]{subfig}% Support for small, `sub' figures and tables

\usepackage[numbers,sort&compress]{natbib}% Citation support using natbib.sty
\usepackage{url}
\bibpunct[, ]{[}{]}{,}{n}{,}{,}% Citation support using natbib.sty
\makeatletter% @ becomes a letter
\def\NAT@def@citea{\def\@citea{\NAT@separator}}% Suppress spaces between citations using natbib.sty
\makeatother% @ becomes a symbol again

\theoremstyle{plain}% Theorem-like structures provided by amsthm.sty
\newtheorem{theorem}{Theorem}[section]
\newtheorem{lemma}[theorem]{Lemma}

\newtheorem{algo}[theorem]{Algorithm}

\theoremstyle{definition}
\newtheorem{definition}[theorem]{Definition}

\theoremstyle{remark}
\newtheorem{remark}{Remark}

\begin{document}

\articletype{RESEARCH ARTICLE}% Specify the article type or omit as appropriate

\title{Standard Curves for Empirical Likelihood Ratio Tests of Means}

\author{
\name{Jost Viebrock\textsuperscript{a} and Thorsten Dickhaus\textsuperscript{b}\thanks{CONTACT Thorsten Dickhaus. Email: dickhaus@uni-bremen.de}}
\affil{\textsuperscript{a}Department of Biometry and Data Management, Leibniz Institute for Prevention Research and Epidemiology – BIPS, Achterstr. 30, 28359 Bremen, Germany; \textsuperscript{b}Institute for Statistics, University of Bremen, Bibliothekstr. 1, 28359 Bremen, Germany}
}

\maketitle

\begin{abstract}
We present simulated standard curves for the calibration of empirical likelihood ratio (ELR) tests of means. With the help of these curves, the nominal significance level of the ELR test can be adjusted in order to achieve (quasi-) exact type I error rate control for a given, finite sample size. By theoretical considerations and by computer simulations, we demonstrate that the adjusted  significance level depends most crucially on the skewness and on the kurtosis of the parent distribution. For practical purposes, we tabulate adjusted critical values under several prototypical statistical models. 
\end{abstract}

\begin{keywords}
Kurtosis; nonparametric test; significance level; skewness; statistical functional; Wilks phenomenon
\end{keywords}

\section{Introduction}\label{sec1}
Testing of statistical functionals is one major topic of nonparametric statistics; see, e.\ g., \cite{dickhaus}. There are several classes of test procedures for testing a (point) null hypothesis about the mean of a probability distribution on $\mathbb{R}^d$ based on a sample of size $n$: (i) For large $n$, one may conduct a $Z$-test based on an appropriate central limit theorem. (ii) Bootstrap tests as introduced in \cite{efron1979annals} are popular in practice. Often, they are asymptotically effective with respect to a $Z$-test, but improve the finite sample properties of the latter regarding the accuracy of type I error control. (iii) Another class of tests is constituted by projection tests in the sense of Chapter 7 in \cite{dickhaus}, in particular (nonparametric) exponential tilting tests (see, among many others, \cite{Csiszar1975} and \cite{Efron1981Canada}) and so-called empirical likelihood ratio (ELR) tests; see \cite{Owen-EL-Book} for an overview. ELR tests can conveniently be carried out as asymptotic chi-square tests, due to a nonparametric "Wilks phenomenon"; cf. \cite{Wilks1938}. However, the resulting chi-square test typically violates the significance level for finite sample sizes. We will provide more details on this issue in Section \ref{sec2} below.

In the present work, we propose a straightforward, simulation-based approach to the calibration of an ELR test. In this, our goal is type I error rate control at (quasi-) exact level $\alpha \in (0, 1)$ in the presence of a given, finite sample size $n$. To this end, we introduce the notion of a "standard curve" in the ELR context. According to \url{https://en.wikipedia.org/wiki/Standard_curve} (latest access: May 5th, 2021), a "standard curve, also known as a calibration curve, is a type of graph used as a quantitative research technique. Multiple samples with known properties are measured and graphed, which then allows the same properties to be determined for unknown samples by interpolation on the graph. The samples with known properties are the standards, and the graph is the standard curve." Our proposal is to take as the "standards" a grid of nominal significance levels for the asymptotic ELR test under a given statistical model. Interpolation of the resulting realized type I error probabilities and solving for $\alpha$ leads to an adjusted nominal significance level $\alpha_{\text{approx}}$ at which the chi-square test can be carried out. For practical purposes, we demonstrate that $\alpha_{\text{approx}}$ most crucially depends on the skewness and the kurtosis of the parent distribution, such that the same value of $\alpha_{\text{approx}}$ may be used for all parent distributions which have (roughly) matching third and fourth (central) moments. The resulting quasi-exact critical value for the ELR test can then simply be looked up from a table.

The remainder of this work is organized as follows. In Section \ref{sec2}, we set up basic notation, and we formalize our problem at hand mathematically. Section \ref{sec3} contains our results with respect to simulated standard curves. In Section \ref{sec4}, quasi-exact critical values for the ELR test of a mean are tabulated under several prototypical models. We conclude with a discussion in Section \ref{sec5}. 

%%%%%%%%%%%%%%%%%%%%%%%%%%%%%%%%%%%%%%%%%%%%%%%
%                S E C T I O N   2            %
%%%%%%%%%%%%%%%%%%%%%%%%%%%%%%%%%%%%%%%%%%%%%%%
\section{Notation and preliminaries}\label{sec2}
Throughout the work, we let $Y_1, \ldots, Y_n$ denote a sample of observable,  $\mathbb{R}^d$-valued, stochastically independent and identically distributed (i.i.d.) random variables, where $n \in \mathbb{N}$ denotes the sample size and $d \in \mathbb{N}$ will be referred to as the dimension. We assume that $Y_1, \ldots, Y_n$ are all defined on the same probability space $(\Omega, \mathcal{F}, \mathbb{P})$. The probability distribution of $Y_1$ will be denoted by $P := \mathbb{P}^{Y_1}$, and we consider the case that $P$ is unknown. Assuming that the mean of $Y_1$ exists in $\mathbb{R}^d$, we are concerned with significance tests for 
\begin{equation}\label{testproblem}
H_0: \mathbb{E}[Y_1]= \mu_0\text{~~versus~~}H_1: \mathbb{E}[Y_1] \neq \mu_0,
\end{equation}
where $\mu_0$ is a given point in $\mathbb{R}^d$.

At least since the work in \cite{owen1990}, one popular class of tests for the test problem \eqref{testproblem} is constituted by ELR tests.

\begin{definition}
Let $\hat{P}_n = n^{-1} \sum_{i=1}^n \delta_{Y_i}$ denote the empirical measure induced by $Y_1, \ldots, Y_n$, where $\delta_a$ denotes the Dirac measure in the point $a$. For given realizations $Y_1(\omega) = y_1, \ldots, Y_n(\omega) = y_n$, the "realized empirical measure" $\hat{P}_n(\omega)$ is a discrete probability measure which puts a point mass of $n^{-1}$ into each of the observation points $y_1, \ldots, y_n$, where we assume for simplicity that there are no ties among the observations.

Furthermore, let $p_1, \ldots, p_n$ be defined by the following constrained optimization problem.
\begin{align}
\text{Maximize}   \quad  &\prod_{i=1}^n n p_i \label{target-cons}\\
\text{subject to} \quad  &\forall 1 \leq i \leq n: p_i \geq 0,\nonumber\\
                         &\sum_{i=1}^n p_i = 1,\nonumber\\ %\label{cons1}\\
                         &\sum_{i=1}^n p_i (y_i - \mu_0) = 0 \in \mathbb{R}^d.\nonumber %\label{cons2}
\end{align}
We denote by $\hat{P}_0(\omega)$ the discrete probability measure which puts the point mass $p_i$ into $y_i$ for all $1 \leq i \leq n$.

Finally, we let
\begin{equation}\label{ratio-mu0}
\mathcal{R}(\mu_0) = \frac{Z(y, \hat{P}_0)}{Z(y, \hat{P}_n)} = \frac{\prod_{i=1}^n p_i}{\prod_{i=1}^n n^{-1}} = \prod_{i=1}^n n p_i,
\end{equation}
where $Z(y, \cdot)$ denotes the nonparametric likelihood function given the data $y = (y_1, \ldots, y_n)^\top$. We call $\mathcal{R}(\mu_0)$ the ELR pertaining to $\mu_0$.
\end{definition}

\begin{remark} $ $
\begin{itemize}
\item[(a)] It is well known that $\hat{P}_n$ is the nonparametric maximum likelihood estimator (NPMLE) of $P$; see, e.\ g., Theorem 2.5 in \cite{dickhaus}. 
\item[(b)] The optimization problem \eqref{target-cons} has a unique solution if $\mu_0$ is located inside the convex hull of $y_1, \ldots, y_n$. In the latter case, $p_1, \ldots, p_n$ can be found by the method of Lagrange multipliers; see, e.\ g., Lemma 7.3 in \cite{dickhaus}.
\end{itemize}
\end{remark}

In the sense of nonparametric likelihood maximization, the ELR $\mathcal{R}(\mu_0)$ defined in \eqref{ratio-mu0} can be interpreted as the goodness-of-fit of the restricted model under $H_0$ relative to the goodness-of-fit in the full (unrestricted) model. Therefore, $\mathcal{R}(\mu_0)$ is a meaningful test statistic for the test problem \eqref{testproblem}. If the resulting ELR test shall be carried out as a significance test at a given significance level $\alpha \in (0, 1)$, the null distribution of $\mathcal{R}(\mu_0)$ or a suitable approximation thereof is required. To this end, the following asymptotic result, the proof of which can be found in \cite{owen1990} and \cite{dickhaus}, is helpful.

\begin{theorem} \label{main-theorem-ELR}
Assume that $Y_1$ possesses the mean $\mu = \mathbb{E}[Y_1] \in \mathbb{R}^d$ as well as a finite and positive definite covariance matrix $\Sigma \in \mathbb{R}^{d \times d}$. Then, the following assertions hold true.
\begin{itemize}
\item[(a)] Under $H_0$, the statistic $- 2 \log\left(\mathcal{R}(\mu_0)\right)$ converges in distribution to $\chi^2_d$ as $n \to \infty$.
\item[(b)] Let $\alpha \in (0, 1)$, and let $c_\alpha = \chi^2_{d; 1 -\alpha}$ denote the $(1 - \alpha)$-quantile of $\chi^2_d$. Then, the set
$$
C_\alpha(\mu) = \left\{\tilde{\mu} = \sum_{i=1}^n p_i Y_i: - 2 \log\left(\mathcal{R}(\tilde{\mu}\right) \leq c_\alpha, \forall 1 \leq i \leq n: p_i \geq 0, \sum_{i=1}^n p_i = 1\right\}
$$
constitutes an asymptotic $(1 - \alpha)$-confidence region for $\mu$, where $n \to \infty$.
\item[(c)] The set $C_\alpha(\mu)$ from part (b) is a convex subset of $\mathbb{R}^d$.
\end{itemize}
\end{theorem}
By duality of tests and confidence regions (see, e.\ g., \cite{Aitchison1964}), $H_0$ can be rejected at asymptotic ($n \to \infty$) significance level $\alpha$, if $C_\alpha(\mu)$ does not cover $\mu_0$ from \eqref{testproblem}, or equivalently, if  $- 2 \log\left(\mathcal{R}(\mu_0)\right)$ exceeds $c_\alpha = \chi^2_{d; 1 -\alpha}$.
Unfortunately, though, the finite sample coverage properties of $C_\alpha(\mu)$ are often not satisfactory, as mentioned for instance in \cite{LaScala1990}, \cite{DiCiccio1991}, \cite{Owen2013}, \cite{Tsao2013}, \cite{Tsao-Wu2013}, \cite{Tsao-Wu2014}, \cite{Wu-Tsao2014}, and \cite{profile-ELR-rho}.
Therefore, we will in the next sections present a Monte Carlo-based method how to adjust the nominal significance level $\alpha$ in the definition of the critical value $c_\alpha$, such that the ELR test becomes an exact level $\alpha$ test in some prototypical model classes.

%%%%%%%%%%%%%%%%%%%%%%%%%%%%%%%%%%%%%%%%%%%%%%%
%                S E C T I O N   3            %
%%%%%%%%%%%%%%%%%%%%%%%%%%%%%%%%%%%%%%%%%%%%%%%
\section{Simulated standard curves}\label{sec3}
In this section, we explain how we have derived standard curves (in the sense mentioned in the introduction) for the calibration of the nominal significance level in the context of ELR tests of means. Under some prototypical statistical models, our (simulated) standard curves will visualize the size (i.\ e., the realized significance level / the realized type I error probability) of the ELR test or equivalently, the realized coverage probability of the confidence region $C_\alpha(\mu)$ from part (b) of Theorem \ref{main-theorem-ELR} as a function of the sample size $n$ or as a function of the nominal significance level $\alpha$, respectively. For simplicity and convenience of the computations, we restrict our study to dimension $d=1$ here. Higher dimensions can in principle be treated analogously.  

First, we recall an important theoretical result.

\begin{theorem}[cf.\ Theorem 3.1 in \cite{zhang1996accuracy}]\label{Zhang-theorem}
Let $d=1$ and assume that $\mathbb{E}[Y_1^8] < \infty$ as well as $\limsup_{|t | \rightarrow \infty}  |\mathbb{E}[\exp (it (Y_1- \mu_0)) ]| < 1$ (Cram\'{e}r's condition). Let $s_1$ denote the skewness and $s_2$ the kurtosis of $Y_1$, and let $\ell(\mu_0) = - 2 \log\left(\mathcal{R}(\mu_0)\right)$. Then, it holds that
\begin{equation}\label{Zhang}
    \mathbb{P}\left(\ell(\mu_0)  \leq c_{\alpha} \right) = 1-\alpha + \frac{1}{2n}\left( \frac{s_2}{2} - \frac{s_1^2}{3}  \right) \cdot  \int_{-\sqrt{c_{\alpha}}}^{\sqrt{c_{\alpha}}}(x^2-1) \phi(x)dx + O(n^{-3/2}),
\end{equation} 
whenever $H_0$ from \eqref{testproblem} is true. In \eqref{Zhang}, $\phi$ denotes the Lebesgue density of the standard normal distribution on $\mathbb{R}$.
\end{theorem}

\begin{lemma}\label{positive-negative}
Under any distribution $P$ fulfilling the assumptions of Theorem \ref{Zhang-theorem}, it necessarily holds that
\begin{equation}\label{positive}
\frac{s_2}{2} > \frac{s_1^2}{3}.
\end{equation}

Furthermore, we have that 
\begin{equation}\label{negative}
\int_{-\sqrt{c_{\alpha}}}^{\sqrt{c_{\alpha}}}(x^2-1) \phi(x)dx < 0
\end{equation}
for all $\alpha \in (0, 1)$.
\end{lemma}

\begin{proof}
To prove \eqref{positive}, we notice that due to the Pearson inequality (see, e.\ g., Equation (1) in \cite{Sharma-Bhandari2015}), it holds that $s_2/2 \geq (s_1^2+1) /2$. However, we have that  $(s_1^2+1) /2 > s_1^2/3$, completing the proof of \eqref{positive}.

To prove \eqref{negative}, we employ integration by parts to see that
\[
\int_{-\sqrt{c_{\alpha}}}^{\sqrt{c_{\alpha}}}(x^2-1) \phi(x)dx = -\sqrt{\frac{2}{\pi}} \sqrt{c_\alpha} \exp\left(- \frac{ c_\alpha}{2} \right).
\]
The proof is completed by noticing that $c_\alpha > 0$.
\end{proof}

The assertions of Theorem \ref{Zhang-theorem} and Lemma \ref{positive-negative} indicate that the size of the ELR test is (for a given nominal significance level $\alpha$) mainly governed by the skewness and by the kurtosis of $Y_1$, at least for larger values of $n$. Furthermore, we may expect that the graphs of our standard curves for the realized size of the ELR test as a function of $n$ will roughly look like a decreasing branch of a hyperbola with limit $\alpha$ as $n \to \infty$. 

The simulation scheme that we have employed to derive the curves in the subsequent sections can be summarized as follows.

\begin{algo} $ $
\begin{itemize}
    \item[(i)] Fix a probability distribution $P$ for $Y_1$, a sample size $n$, a nominal significance level $\alpha$, and a number $B$ of Monte Carlo repetitions. 
    \item[(ii)] For $b$ from $1$ to $B$ do:
    \begin{itemize}
        \item[(a)] Generate an i.i.d. (pseudo) sample $y_1^{(b)}, \ldots, y_n^{(b)}$ from $P$ on the computer.
        \item[(b)] Compute $\ell^{(b)}(\mu_0)$ from $y_1^{(b)}, \ldots, y_n^{(b)}$, where $\mu_0$ is the true expected value of $Y_1$ under $P$ and the notation $\ell^{(b)}(\mu_0)$ is as in Theorem \ref{Zhang-theorem}, where the superscript refers to the simulation run.
    \end{itemize}
    \item[(iii)] Let $\hat{\alpha}_n = B^{-1} \sum_{b=1}^B \mathbf{1}\{\ell^{(b)}(\mu_0) > c_\alpha\}$, and take $\hat{\alpha}_n$ as the approximation of the realized type I error probability of the ELR test for $\mu_0$ for the chosen $P$, $n$, and $\alpha$.
\end{itemize}
\end{algo}

Throughout the remainder, we will consider the classes of parent distributions for $Y_1$ which are listed in Table \ref{parent-distributions}.

\begin{table}[htp]
\tbl{Classes of parent distributions for $Y_1$ which have been considered in our simulations.}
{\begin{tabular}{lccc} \toprule
 Class of distributions &Notation and parameters &Support &Lebesgue density\\
\midrule
Normal & $\mathcal{N}(\mu,\sigma^2)$& $\mathbb{R}$ & $f_{\mu,\sigma^2}(y) = \frac{1}{\sqrt{2 \pi \sigma^2}} \exp \left( - \frac{(y- \mu)^2}{2 \sigma^2} \right) $ \\
Exponential & Exp($\lambda$) & $\mathbb{R}_{>0}$ & $f_\lambda(y)= \lambda \exp (-\lambda y)$ \\
Uniform &Unif$(a,b)$ & $[a,b]$ & $f_{a,b}(y)= \frac{1}{b-a}$ \\
Gamma & $\Gamma(\alpha,\beta)$ & $\mathbb{R}_{>0}$& $f_{\alpha,\beta}(y) = \frac{\beta^\alpha}{\Gamma(\alpha)}x^{\alpha-1} \exp (- \beta y )$ \\
Chi-square & $ \chi^2(\nu)$ or $\chi^2_\nu$ & $\mathbb{R}_{>0}$ & $f_\nu(y) = \frac{1}{2^{\nu/2} \Gamma(\nu/2)} y^{2/\nu-1} \exp (-y/2 )$
\\
Laplace & Lap$(\mu, b)$ & $\mathbb{R}$ & $f_{\mu,b}(y)= \frac{1}{2b} \exp \left( - \frac{|y-\mu|}{b} \right)$\\
Student's $t$ & $t(\nu)$ & $\mathbb{R}$ & $f_\nu(y)=\frac{\Gamma((\nu+1)/2)}{\sqrt{\nu \pi} \Gamma(\nu/2)} \left( 1+ \frac{y^2}{\nu} \right)^{-\frac{\nu+1}{2}} $\\ 
\bottomrule
\end{tabular}}
\tabnote{Outside of their supports, the Lebesgue densities may be set to zero. The symbol $\Gamma$ denotes the gamma function with $\Gamma(z) = \int_0^\infty x^{z-1} \exp (-x) dx$.}
\label{parent-distributions}
\end{table}

%%%%%%%%%%%%%%%%%%%%%%%%%%%%%%%%%%%%%%%%%%%%%%%%%%%%%%%%%%%%%%%%%%
\subsection{Standard curves for fixed significance level $\alpha$ as a function of the sample size $n$}

\begin{figure}[htp]
\centering
\includegraphics[width=\textwidth]{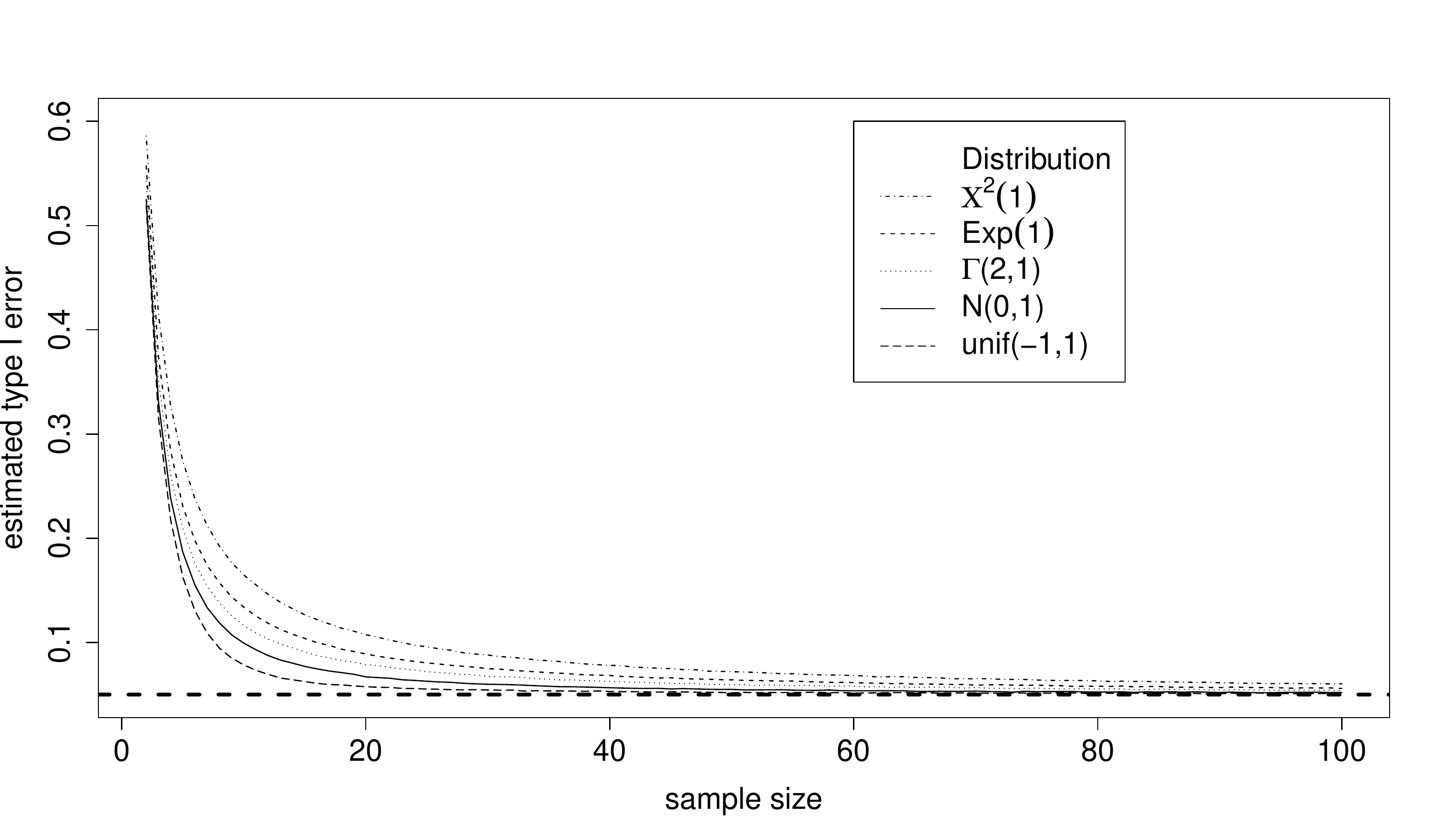}
\caption{Standard curves for the ELR test of a univariate mean. The nominal significance level $\alpha =0.05$ is indicated by the dashed horiziontal line. The number of Monte Carlo repetitions has been set to $B = 10^6$.} \label{thesis-fig4_2}
\end{figure}

Figure \ref{thesis-fig4_2} displays standard curves for five probability distributions from different classes appearing in Table \ref{parent-distributions}. The nominal significance level has been set to $\alpha = 5\%$ and is indicated by the dashed horizontal line. The number of Monte Carlo repetitions has been set to $B = 10^6$.

%%%%%%%%%%%%%%%%%%%%%%%%
% Five standard curves %
%%%%%%%%%%%%%%%%%%%%%%%%

\begin{figure}[htp]
\centering
\includegraphics[width=\textwidth]{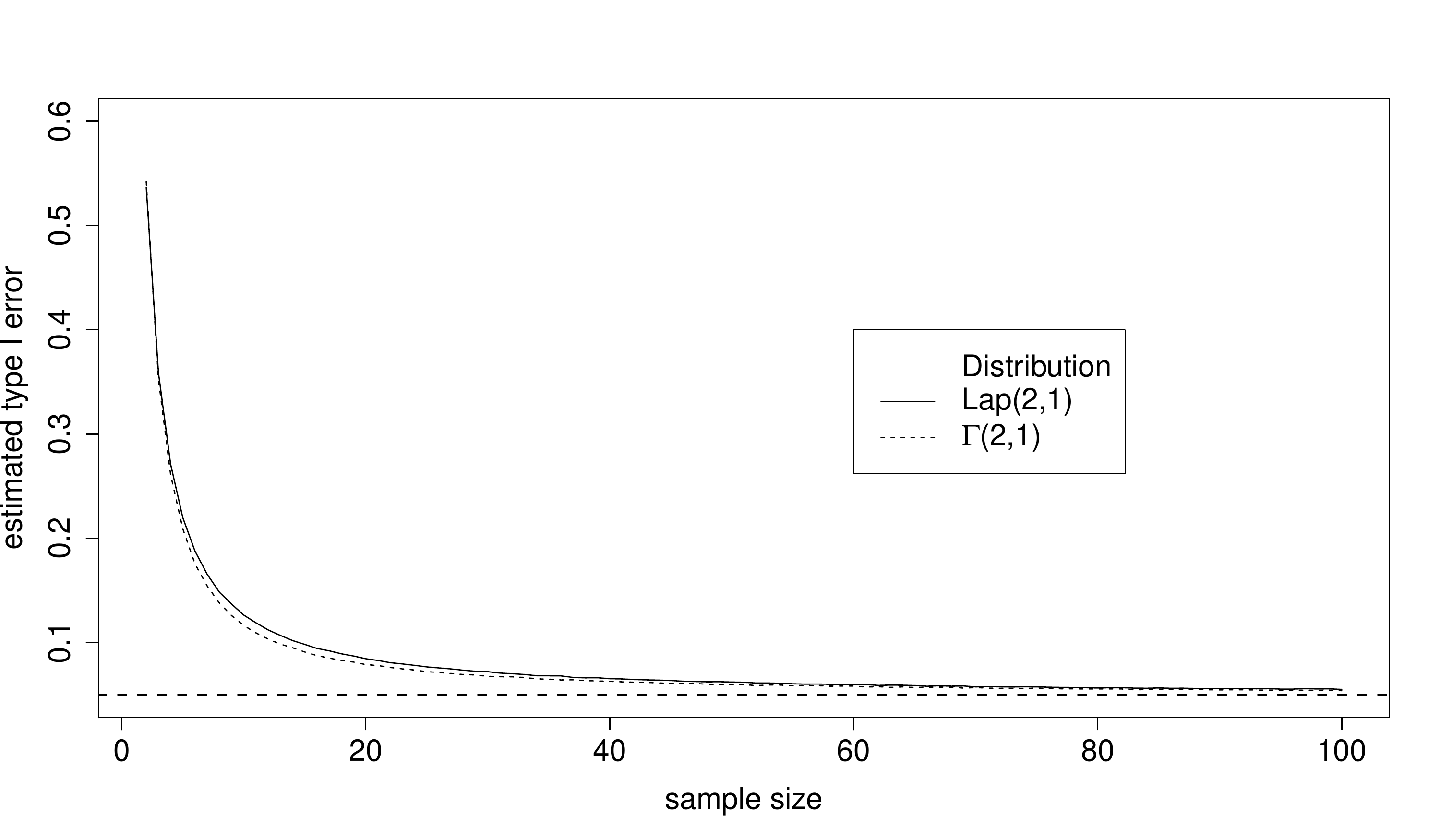}
\caption{Standard curves for the ELR test of a univariate mean. The nominal significance level $\alpha =0.05$ is indicated by the dashed horiziontal line. The number of Monte Carlo repetitions has been set to $B = 10^6$. The two parent distributions indicated in the legend have the same mean, the same variance, and the same kurtosis, but different skewnesses.} \label{thesis-fig4_4}
\end{figure}

\begin{figure}[htp]
\centering
\includegraphics[width=\textwidth]{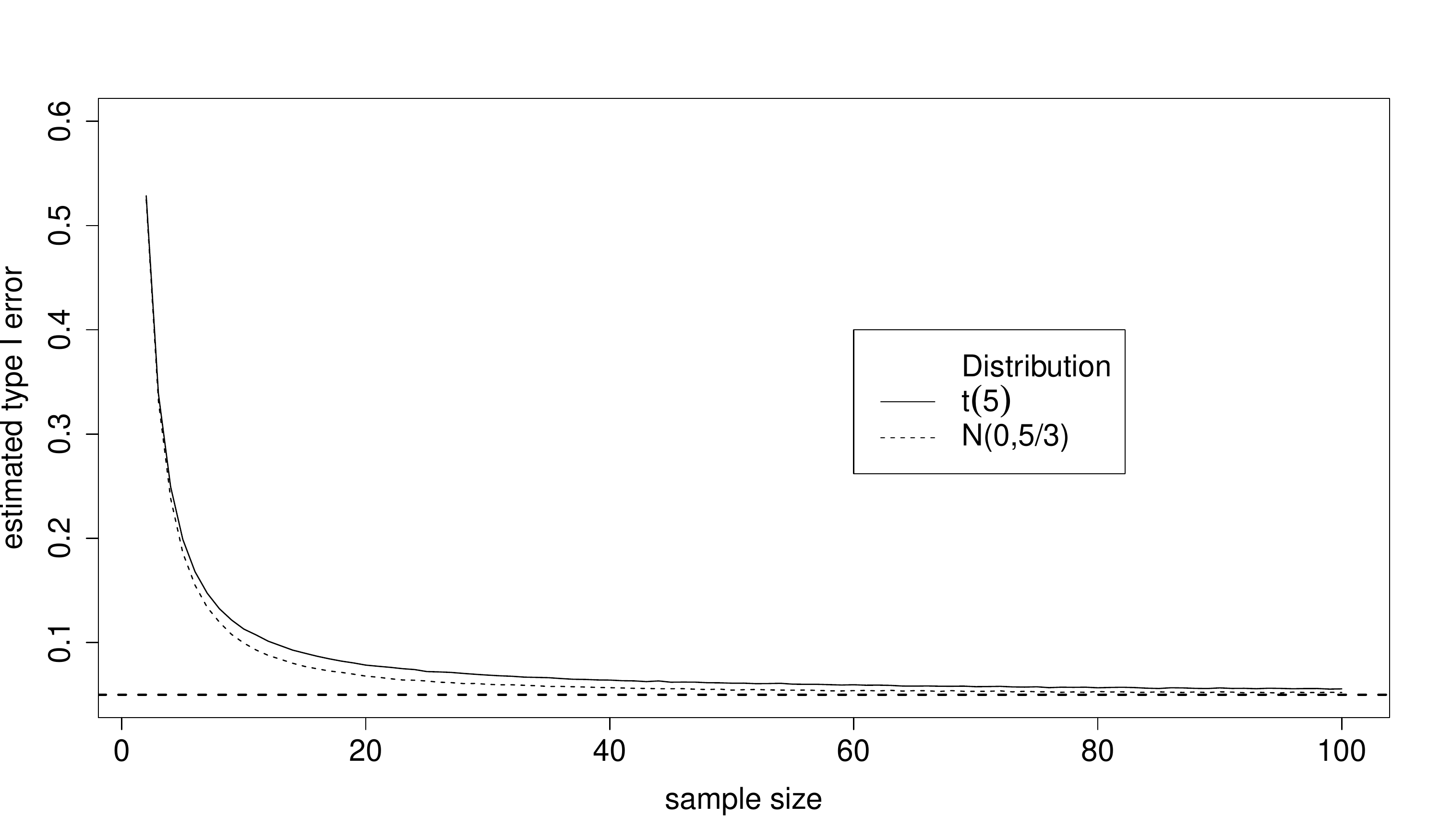}
\caption{Standard curves for the ELR test of a univariate mean. The nominal significance level $\alpha =0.05$ is indicated by the dashed horiziontal line. The number of Monte Carlo repetitions has been set to $B = 10^6$. The two parent distributions indicated in the legend have the same mean, the same variance, and the same skewness, but different kurtoses.} \label{thesis-fig4_5}
\end{figure}

The general shape of the curves (resembling a decreasing branch of a hyperbola) in Figure \ref{thesis-fig4_2} is as expected from the  assertions of Theorem \ref{Zhang-theorem} and Lemma \ref{positive-negative}. The exact values of the $\hat{\alpha}_n$'s seem to depend on the skewness and the kurtosis pertaining to $P$, which is in line with \eqref{Zhang}. To substantiate the latter point further, we compare in Figure \ref{thesis-fig4_4} two parent distributions with the same mean, variance, and kurtosis, but different skewness, and in Figure \ref{thesis-fig4_5} we compare two parent distributions with the same mean, variance, and skewness, but different kurtosis.

%%%%%%%%%%%%%%%%%%%%%%%%%%
% Comparison of skewness %
%%%%%%%%%%%%%%%%%%%%%%%%%%

In Figure \ref{thesis-fig4_4}, the curve pertaining to the (skewed) gamma distribution decreases faster than the curve pertaining to the (symmetric) Laplace distribution. This behavior is in line with the assertion of Theorem \ref{Zhang-theorem}, because in \eqref{Zhang} the term involving $s_1^2$ has a negative sign. Hence, for fixed mean, variance, and kurtosis, a larger squared skewness should lead to a somewhat better approximation quality of the exact $(1-\alpha)$-quantile of the null distribution of $\ell(\mu_0)$ by the asymptotic chi-square quantile $c_\alpha$.  

%%%%%%%%%%%%%%%%%%%%%%%%%%
% Comparison of kurtosis %
%%%%%%%%%%%%%%%%%%%%%%%%%%

Analogously, in Figure \ref{thesis-fig4_5} the curve pertaining to the normal distribution decreases faster than the curve pertaining to Student's $t$-distribution with five degrees of freedom. This is also in line with the assertion of Theorem \ref{Zhang-theorem}, because in \eqref{Zhang} the term involving $s_2$ has a positive sign. Hence, for fixed mean, variance, and skewness, a larger kurtosis should lead to a somewhat worse approximation quality. And it is well-known that the $t(5)$-distribution has a positive excess kurtosis, meaning that its kurtosis is larger than $3$, which is the kurtosis of a normal distribution on $\mathbb{R}$.

\begin{figure}[htp]
\centering
\includegraphics[width=\textwidth]{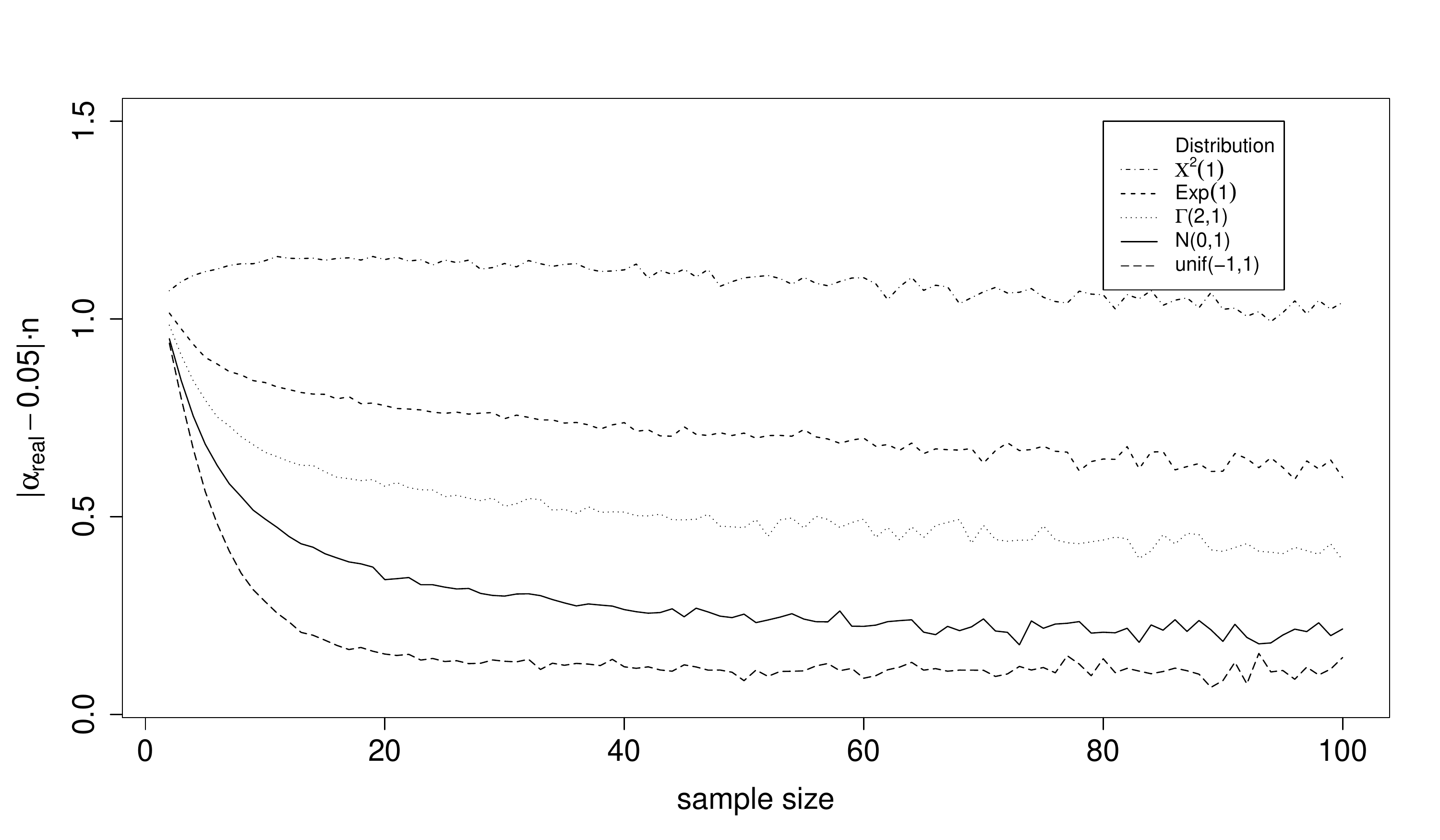}
\caption{Some values of $n \cdot |\hat{\alpha}_n - \alpha|$ for several parent distributions $P$. The simulation settings are as in Figure \ref{thesis-fig4_2}.} \label{thesis-fig4_6}
\end{figure}

Finally, we assess the impact of the $O(n^{-3/2})$ terms appearing in \eqref{Zhang} for finite sample sizes. To this end, we plot in Figure \ref{thesis-fig4_6} values of $n \cdot |\hat{\alpha}_n - \alpha|$ for several parent distributions $P$. In the case that the $O(n^{-1})$ terms dominate the $O(n^{-3/2})$ terms already in the cases of moderate sample sizes, the scaled absolute differences $\{n \cdot |\hat{\alpha}_n - \alpha|\}_{n > 1}$ should stabilize for increasing $n$. This effect can clearly be observed in Figure \ref{thesis-fig4_6}.

%%%%%%%%%%%%%%%%%%%%%%%%%%%%%%%%%%%%%%%%%%%%%%%%%%
%       Impact of O(n^{-3/2}) terms              %
%%%%%%%%%%%%%%%%%%%%%%%%%%%%%%%%%%%%%%%%%%%%%%%%%%

%%%%%%%%%%%%%%%%%%%%%%%%%%%%%%%%%%%%%%%%%%%%%%%%%%%%%%%%%%%%%%%%%%
\subsection{Standard curves for fixed sample size $n$ as a function of the significance level $\alpha$}
\begin{figure}[htp]
\centering
\includegraphics[width=\textwidth]{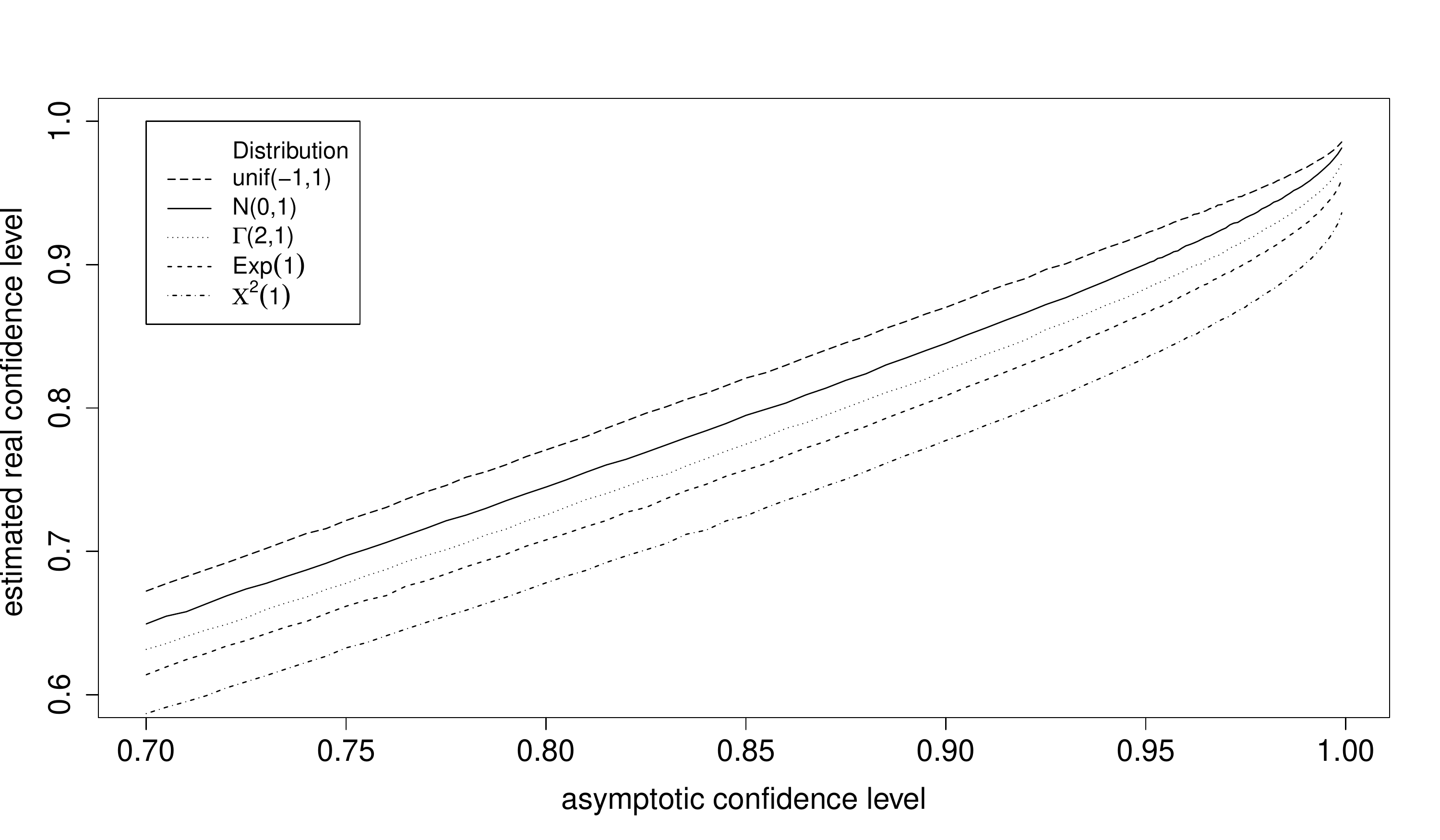}
\caption{Standard curves for the ELR test of a univariate mean. The sample size is fixed at $n=10$. The number of Monte Carlo repetitions has been set to $B = 10^6$.} \label{thesis-fig4_9}
\end{figure}

From a practical point of view, the sample size $n$ is often fixed for the experiment at hand. In such a case, it is more informative to plot $\hat{\alpha}_n$ against $\alpha$ or $1 - \hat{\alpha}_n$ against $1 - \alpha$, respectively. For the specific choice of $n = 10$, such curves are provided in Figure \ref{thesis-fig4_9}. 

As we will demonstrate in Section \ref{sec4}, standard curves as in Figure \ref{thesis-fig4_9} are helpful for determining a quasi-exact critical value for the ELR test for a given, finite sample size. Namely, one may for a fixed target confidence level $1-\alpha$ and for a certain $P$ find the point on the respective standard curve with ordinate $1-\alpha$, and use the corresponding abscissa value (which will typically be larger than $1-\alpha$) as the nominal confidence level for constructing the confidence set for $\mu$. For example, if $n=10$ and $1 - \alpha = 0.8$ for an assumed symmetric and (approximately, at least in terms of the kurtosis of $Y_1$) uniform distribution under the null, one should use as the nominal confidence level a value of approximately $0.85$ based on the uppermost standard curve displayed in Figure \ref{thesis-fig4_9}. On the contrary, if the assumed kurtosis of $Y_1$ is assumed to be close to $3$ as in the case of the normal distribution, the solid standard curve displayed in Figure \ref{thesis-fig4_9} suggests to take a slightly larger nominal confidence level of approximately $0.88$.

%%%%%%%%%%%%%%%%%%%%%%%%%%%%%%%%%%%%%%%%%%%%%%%
%                S E C T I O N   4            %
%%%%%%%%%%%%%%%%%%%%%%%%%%%%%%%%%%%%%%%%%%%%%%%
\section{Simulated critical values}\label{sec4}
In this section, we elaborate further on the quasi-exact calibration of the critical value for the ELR test for $\mu$. We have already outlined our proposed standard curve-based approach for this in our discussion of  Figure \ref{thesis-fig4_9}. 

Formally, assume that a target significance level $\alpha \in (0, 1)$ for the ELR test for $\mu$ is given, and that the sample size $n$ is fixed. Furthermore, assume that a standard curve as in Figure \ref{thesis-fig4_9} is at hand, where the "standards" refer to several values of $\hat{\alpha}_n$ which have been included in the computation of the curve.

Define now
\begin{align*}
   \hat{\alpha}_n^{(1)} := \max\big\{ \hat{\alpha}_n \mid \hat{\alpha}_n \leq \alpha\big\}  \quad \text{~~and~~} \quad
   \hat{\alpha}_n^{(2)} :=
   \min \big\{ \hat{\alpha}_n \mid \hat{\alpha}_n \geq \alpha\big\}.
\end{align*} 

Our proposal is to (linearly) interpolate between $\hat{\alpha}_n^{(1)}$ and $\hat{\alpha}_n^{(2)}$, find the resulting standard curve abscissa $\alpha_{\text{approx}}$ (say), and use as the quasi-exact critical value for the ELR test for $\mu$ the $(1-\alpha_{\text{approx}})$-quantile of the chi-square distribution with $d=1$ degrees of freedom (in the case of a test for a univariate mean). This critical value will be denoted by $\hat{c}_\alpha = F_{\chi^2_1}^{-1}\left(1-\alpha_{\text{approx}}\right)$, where $F_{\chi^2_1}$ denotes the cumulative distribution function of the chi-square distribution with one degree of freedom. Tables \ref{critical-values-normal} - \ref{critical-values-chisq} display the resulting quasi-exact critical values for several parent distributions $P$. Cell entries \texttt{NA} indicate parameter constellations for which the solution could not be obtained with our method, due to numerical issues. Such issues have occurred for small sample sizes and/or small nominal significance levels.

\begin{remark} $ $
\begin{itemize}
\item[(a)] For uniqueness of $\alpha_{\text{approx}}$, the considered standard curve has to be injective. This property was fulfilled in all our simulations.
\item[(b)] Utilizing $B = 10^6$ Monte Carlo repetitions in the simulations of the standard curves implies that the Monte Carlo error of $\hat{\alpha}_n$ is of the order of magnitude of $10^{-4}$. Hence, one may decrease  $\alpha_{\text{approx}}$ by, e.\ g., $5 \times 10^{-4}$ if a conservative type I error behavior of the ELR test is desired. This may also robustify the standard curve-based method against the impact of (central) moments of order higher than four. See, for example, Section 4 in \cite{MVCHS} for a more detailed discussion about the required number of Monte Carlo simulations for a given error tolerance.
\end{itemize}
\end{remark}

\begin{table}[htp]
\tbl{Simulated quasi-exact critical values for the ELR test of the mean of a normal distribution}
{\begin{tabular}{l|ccccccccc} \toprule
 $n$ \textbackslash $1-\alpha$ & 0.7 & 0.8 & 0.85 & 0.9 & 0.95 & 0.96 & 0.97 & 0.98 & 0.99\\
\midrule
10 & 1.342 & 2.134 & 2.777 & 3.831 & 6.054 & 6.953 & 8.183 & 10.285 & NA \\
15 & 1.226 & 1.905 & 2.444 & 3.265 & 4.873 & 5.469 & 6.267 & 7.444 & 9.82 \\
20 & 1.181 & 1.824 & 2.321 & 3.076 & 4.491 & 4.986 & 5.624 & 6.607 & 8.461 \\
30 & 1.138 & 1.75 & 2.216 & 2.911 & 4.183 & 4.612 & 5.204 & 6.017 & 7.54 \\
50 & 1.11 & 1.699 & 2.147 & 2.812 & 4.01 & 4.416 & 4.925 & 5.712 & 7.046 \\
100 & 1.093 & 1.666 & 2.11 & 2.752 & 3.902 & 4.307 & 4.801 & 5.502 & 6.785\\
\bottomrule
\end{tabular}}
\tabnote{Notice that all normal distributions on $\mathbb{R}$ have a skewness of zero and a kurtosis of $3$.}
\label{critical-values-normal}
\end{table}

\begin{table}[htp]
\tbl{Simulated quasi-exact critical values for the ELR test of the mean of an exponential distribution}
{\begin{tabular}{l|ccccccccc} \toprule
 $n$ \textbackslash $1-\alpha$ & 0.7 & 0.8 & 0.85 & 0.9 & 0.95 & 0.96 & 0.97 & 0.98 & 0.99\\
\midrule
10 & 1.584 & 2.58 & 3.442 & 4.976 & 9.019 & NA & NA & NA & NA \\
15 & 1.389 & 2.202 & 2.884 & 3.953 & 6.243 & 7.163 & 8.496 & 10.76 & NA \\
20 & 1.303 & 2.046 & 2.639 & 3.558 & 5.414 & 6.105 & 7.055 & 8.552 & NA \\
30 & 1.221 & 1.897 & 2.429 & 3.224 & 4.758 & 5.301 & 6.042 & 7.133 & 9.283 \\
50 & 1.158 & 1.789 & 2.266 & 2.977 & 4.224 & 4.801 & 5.39 & 6.303 & 7.91 \\
100 & 1.115 & 1.71 & 2.164 & 2.836 & 4.056 & 4.474 & 5.007 & 5.79 & 7.156\\
\bottomrule
\end{tabular}}
\tabnote{Notice that all exponential distributions on $\mathbb{R}_{\geq 0}$ have a skewness of $2$ and a kurtosis of $9$.}
\label{critical-values-exponential}
\end{table}

\begin{table}[htp]
\tbl{Simulated quasi-exact critical values for the ELR test of the mean of a uniform distribution}
{\begin{tabular}{l|ccccccccc} \toprule
 $n$ \textbackslash $1-\alpha$ & 0.7 & 0.8 & 0.85 & 0.9 & 0.95 & 0.96 & 0.97 & 0.98 & 0.99\\
\midrule
10 & 1.207 & 1.875 & 2.418 & 3.265 & 5.093 & 5.831 & 6.913 & 8.805 & NA \\
15 & 1.152 & 1.768 & 2.242 & 2.955 & 4.281 & 4.742 & 5.402 & 6.399 & 8.35 \\
20 & 1.126 & 1.734 & 2.186 & 2.873 & 4.11 & 4.519 & 5.093 & 5.894 & 7.375 \\
30 & 1.111 & 1.698 & 2.144 & 2.81 & 3.978 & 4.367 & 4.884 & 5.645 & 6.921 \\
50 & 1.097 & 1.671 & 2.114 & 2.758 & 3.921 & 4.308 & 4.81 & 5.525 & 6.785 \\
100 & 1.083 & 1.655 & 2.094 & 2.734 & 3.891 & 4.259 & 4.757 & 5.453 & 6.696\\
\bottomrule
\end{tabular}}
\tabnote{Notice that all uniform distributions on an intervall $[a, b] \subset \mathbb{R}$ have a skewness of zero and a kurtosis of $9/5$.}
\label{critical-values-uniform}
\end{table}

\begin{table}[htp]
\tbl{Simulated quasi-exact critical values for the ELR test of the mean of the $\Gamma (2,1)$-distribution}
{\begin{tabular}{l|ccccccccc} \toprule
 $n$ \textbackslash $1-\alpha$ & 0.7 & 0.8 & 0.85 & 0.9 & 0.95 & 0.96 & 0.97 & 0.98 & 0.99\\
\midrule
10 & 1.463 & 2.35 & 3.099 & 4.347 & 7.301 & 8.61 & 10.595 & NA & NA \\
15 & 1.312 & 2.058 & 2.656 & 3.597 & 5.508 & 6.234 & 7.275 & 8.905 & NA\\
20 & 1.241 & 1.94 & 2.481 & 3.312 & 4.935 & 5.531 & 6.327 & 7.547 & 9.988  \\
30 & 1.177 & 1.826 & 2.334 & 3.067 & 4.495 & 4.96 & 5.606 & 6.591 & 8.391 \\
50 & 1.133 & 1.743 & 2.21 & 2.891 & 4.184 & 4.609 & 5.181 & 6.014 & 7.47 \\
100 & 1.1 & 1.691 & 2.133 & 2.789 & 3.979 & 4.368 & 4.883 & 5.66 & 6.958 \\
\bottomrule
\end{tabular}}
\label{critical-values-gamma}
\end{table}

\begin{table}[htp]
\tbl{Simulated quasi-exact critical values for the ELR test of the mean of the chi-square distribution with one degree of freedom}
{\begin{tabular}{l|ccccccccc} \toprule
 $n$ \textbackslash $1-\alpha$ & 0.7 & 0.8 & 0.85 & 0.9 & 0.95 & 0.96 & 0.97 & 0.98 & 0.99\\
\midrule
10 & 1.829 & 3.086 & 4.26 & 6.563 & NA & NA & NA & NA & NA \\
15 & 1.547 & 2.504 & 3.298 & 4.69 & 7.978 & 9.451 & NA & NA & NA \\
20 & 1.419 & 2.258 & 2.94 & 4.054 & 6.428 & 7.385 & 8.74 & NA & NA \\
30 & 1.295 & 2.033 & 2.626 & 3.519 & 5.342 & 5.983 & 6.917 & 8.313 & NA  \\
50 & 1.205 & 1.869 & 2.38 & 3.175 & 4.636 & 5.165 & 5.836 & 6.86 & 8.854 \\
100 & 1.135 & 1.746 & 2.217 & 2.915 & 4.191 & 4.62 & 5.205 & 6.052 & 7.549 \\
\bottomrule
\end{tabular}}
\label{critical-values-chisq}
\end{table}

For a comparison, Table \ref{chisq1-quantiles} tabulates quantiles of the (limiting) chi-square distribution with one degree of freedom. 

\begin{table}[htp]
\tbl{Some $(1- \alpha)$-quantiles of the chi-square distribution with one degree of freedom}
{\begin{tabular}{|c|ccccccccc} \toprule
$1-\alpha$ & 0.7 & 0.8 & 0.85 & 0.9 & 0.95 & 0.96 & 0.97 & 0.98 & 0.99 \\
\midrule
$F_{\chi^2_1}^{-1}\left(1-\alpha\right)$ & 1.074 & 1.642 & 2.072 & 2.706 & 3.841 & 4.218 & 4.709 & 5.412 & 6.635 \\
\bottomrule
\end{tabular}}
\label{chisq1-quantiles}
\end{table}

%%%%%%%%%%%%%%%%%%%%%%%%%%%%%%%%%%%%%%%%%%%%%%%
%                S E C T I O N   5            %
%%%%%%%%%%%%%%%%%%%%%%%%%%%%%%%%%%%%%%%%%%%%%%%
\section{Discussion}\label{sec5}
We have introduced the notion of a standard curve for ELR tests of means, and we have demonstrated how standard curves can help to calibrate the exact critical value of such a test. Worksheets in \texttt{R}, with which all results reported in this work can be reproduced, are available from the first author upon request.

From a theoretical perspective, our results shine a light on the relevance of the asymptotic result of Theorem 3.1 in \cite{zhang1996accuracy} for finite sample size $n$. Our simulation results in Section \ref{sec3} indicate that this relevance is rather high and that the $O(n^{-3/2})$ terms appearing in \eqref{Zhang} are often rather negligible, already for small to moderate $n$. From a more application-oriented perspective, our proposed standard curve-based calibration approach for the critical value of the ELR test presented in Section \ref{sec4} may be considered a straightforward alternative to other techniques like Bartlett correction, "extended empirical likelihood" or "adjusted empirical likelihood" (see \cite{AEL}), provided that information about skewness and kurtosis of the parent distribution is available or the sample allows for a reliable estimation of the latter quantities. 

There are several possible extensions of the present work: First, as already mentioned in Section \ref{sec3}, one may consider higher dimensions $d >1$ and study the impact of moments of higher order in this setting. Second, it may be of interest to study the case of dependent observations which constitute a time series. Third, one may consider more general functionals of $P$ than just its mean. The result of Theorem 3.1 in \cite{zhang1996accuracy} is actually derived for the rather general class of $M$-functionals. Finally, the standard curve technique is not limited to the ELR context, but may be applied to other methods of asymptotic statistical inference, too. We reserve these topics for future research.

%%%%%%%%%%%%%%%%%%%%%%%%%%%%%%%%%
%         Backmatter            %
%%%%%%%%%%%%%%%%%%%%%%%%%%%%%%%%%
\section*{Acknowledgements}
We thank Prof. Pierpaolo Brutti for fruitful discussions and for hosting the first author at Sapienza University of Rome. 

\section*{Disclosure statement}
No conflict of interest.

%\section*{Funding}
%An unnumbered section, e.g.\ \verb"\section*{Funding}", may be used for %grant details, etc.\ if required and included \emph{in the non-anonymous %version} before any Notes or References.

%\section*{Notes on contributor(s)}
%An unnumbered section, e.g.\ \verb"\section*{Notes on contributors}", %may be included \emph{in the non-anonymous version} if required. A photograph may be added if requested.

%\section*{Nomenclature/Notation}
%An unnumbered section, e.g.\ \verb"\section*{Nomenclature}" (or %\verb"\section*{Notation}"), may be included if required, before any %Notes or References.

%\section*{Notes}
%An unnumbered `Notes' section may be included before the References (if %using the \verb"endnotes" package, use the command \verb"\theendnotes" %where the notes are to appear, instead of creating a \verb"\section*").

%%%%%%%%%%%%%%%%%%%%%%%%%%%%%%%%%%%%
%           REFERENCES             %
%%%%%%%%%%%%%%%%%%%%%%%%%%%%%%%%%%%%
%\bibliographystyle{tfnlm}
%\bibliography{bibliography}

%\appendix
%\section{This is the title of the first appendix}
%\section{This is the title of the second appendix}
%
\end{document}